  \documentclass[review, number, sort&compress, 12pt]{elsarticle}

	\usepackage{url}

 \usepackage{graphicx}

\usepackage{amssymb,amsbsy}

\usepackage{textcomp}
\usepackage{upgreek}

\journal{NIM A}

\begin{document}

\begin{frontmatter}



\title{A multilayer surface detector for ultracold neutrons}


\author[LANL]{Zhehui Wang\corref{cor1}}
\author[LANL]{M. A. Hoffbauer}
\author[LANL]{C. L. Morris}%
\author[IU]{N. B. Callahan}
\author[IU]{E. R. Adamek}
\author[LANL]{J. D. Bacon}
\author[CSU]{M. Blatnik}
\author[NCSU] {A. E. Brandt}
\author[LANL]{L. J. Broussard}
\author[LANL]{S. M. Clayton}
\author[NCSU]{C. Cude-Woods}
\author[LANL]{S. Currie}
\author[NCSU]{E. B. Dees}
\author[VA]{X. Ding}
\author[LANL]{J. Gao}
\author[Regis]{F. E. Gray}
\author[UCLA]{K. P. Hickerson}
\author[TTU]{A. T. Holley}
\author[LANL]{T. M. Ito}
\author[IU]{C.-Y. Liu}
\author[LANL]{M. Makela}
\author[LANL]{J. C. Ramsey}
\author[LANL]{R. W. Pattie, Jr.}
\author[LANL,IU]{D. J. Salvat}
\author[LANL]{A. Saunders}
\author[LANL]{D. W. Schmidt}
\author[LANL]{R. K. Schulze}
\author[LANL]{S. J. Seestrom}
\author[Russia]{E. I. Sharapov}
\author[UK]{A. Sprow}
\author[LANL]{Z. Tang}
\author[LANL]{W. Wei}
\author[NCSU]{J. Wexler}
\author[LANL]{T. L. Womack}
\author[NCSU]{A. R. Young}
\author[NCSU]{B. A. Zeck}
\cortext[cor1]{Correspondence: zwang@lanl.gov}
\address[LANL]{Los Alamos National Laboratory, Los Alamos, NM 87545, USA}
\address[IU]{Indiana University, Bloomington, IN 47405, USA}
\address[CSU]{Cleveland State University, Cleveland, OH 44115, USA}
\address[NCSU]{North Carolina State University, Raleigh, NC 27695, USA}
\address[VA]{Virginia Polytechnic Institute and State University, Blacksburg, VA 24061, USA}
\address[Regis]{Regis University, Denver, CO 80221, USA}
\address[UCLA]{University of California Los Angeles, Los Angeles, CA 90095, USA}
\address[TTU]{Tennessee Technological University, Cookeville, TN 38505, USA}
\address[Russia]{Joint Institute for Nuclear Research, 141980, Dubna, Russia}
\address[UK]{University of Kentucky, Lexington, KY 40506, USA}

\begin{abstract}
A multilayer surface detector for ultracold neutrons (UCNs) is described. The top $^{10}$B layer is exposed to vacuum and directly captures UCNs. The ZnS:Ag layer beneath the $^{10}$B layer is a few microns thick, which is sufficient to detect the charged particles from the $^{10}$B(n,$\alpha$)$^7$Li neutron-capture reaction, while thin enough that ample light due to $\alpha$ and $^7$Li escapes for detection by photomultiplier tubes. A 100-nm thick $^{10}$B layer gives high UCN detection efficiency, as determined by the mean UCN kinetic energy, detector materials and others. Low background, including negligible sensitivity to ambient neutrons, has also been verified through pulse-shape analysis and  comparisons with other existing $^3$He and $^{10}$B detectors.  This  type of detector has been configured in different ways for UCN flux monitoring, development of UCN guides and neutron lifetime research.
\newline
\end{abstract}


\end{frontmatter}



\newpage
\section{Introduction}
\label{sec:1}

Detection of ultracold neutrons (UCNs), or neutrons with kinetic energies less than about 300 neV (1 neV = 10$^{-9}$ eV), is much like detection of thermal neutrons. That is, the same neutron-capture reactions,  such as $^3$He(n, p)$^3$H, $^6$Li(n,$\alpha$)$^3$H, $^{10}$B(n, $\alpha$)$^7$Li  and $^{157}$Gd(n,$\gamma$)$^{158}$Gd, are used to turn neutrons into charged particles or $\gamma$-rays~\cite{Stedman:1960,CF:1983,Knoll:2000,Morris:1999,Salvat:2012,Jenke:2013}. The charged particles and  $\gamma$-rays released from the capture reactions have kinetic energies ranging from hundreds of keV to a few MeV and can be readily detected using gas ionization chambers or scintillators. 

Detection of UCNs, unlike detecting thermal neutrons, is sensitive to the surface conditions, gravity, magnetic fields and ambient gas conditions. All of these factors can modify UCN velocities and therefore alter inelastic scatterings of UCNs as well as UCN capture or absorption. Sensitivity of UCNs to gravity, magnetic fields, material structures and phases of matter provides opportunities to probe these forces or material structures with a precision that is inaccessible to methods using charged particles. On the other hand, it is a common UCN detector challenge to reduce non-UCN background in all of the measurements since a.) UCN counting rates are typically low and the UCN signals are similar to background signals, in particular background neutron signals that can come from upscattered UCNs, thermal and higher energy neutrons; b.) Production of UCNs  using either nuclear reactors or accelerators also generates higher energy neutrons and $\gamma$-rays that  easily outnumber the UCN population.

The UCN absorption mean free path ($\lambda_a$) is given by~\cite{LS:1965, Ignatovich:1990,Golub:1991}
\begin{equation}
\lambda_a = \tau_av_n,
\label{eq:mfp}
\end{equation} 
where the neutron absorption time ($\tau_a$) in solid $^{10}$B is independent of the neutron velocity ($v_n$) and can be calculated from thermal neutron absorption cross section, $\sigma_{th} =$ 3842 barn for the (n,$\alpha$) process, $\tau_a = n_0\sigma_{th} v_{th} $ = 9.0 ns. Here $n_0$ is the solid density of $^{10}$B and $v_{th}$ the neutron thermal velocity. For UCNs at 4.4 m/s, $\lambda_a = 40$ nm. The de Broglie wavelength of UCN ($\lambda_n$) is longer than $\lambda_a$~\cite{Squires:1978}, 
\begin{equation}
\lambda_n =\frac{904.5}{\sqrt{E_n}} = \frac{395.6}{v_n},
\label{wvleng}
\end{equation}
where $\lambda_n$ is in nm, the kinetic energy of the neutron ($E_n$) in neV and the velocity of the neutron ($v_n$) in m/s. For a UCN with a kinetic energy of 100 neV or a velocity of 4.4 m/s, $\lambda_n = 90$ nm. 

We describe a multilayer surface detector for UCNs based on $^{10}$B thin-film capture of neutrons. The top $^{10}$B layer is exposed to vacuum and directly captures UCNs. The ZnS:Ag luminescent layer is beneath the $^{10}$B layer. The effective ZnS layer thickness measured using a $^{148}$Gd $\alpha$ source is a few microns thick, which is sufficient to stop the charges from the $^{10}$B(n,$\alpha$)$^7$Li neutron-capture reaction while thin enough that light due to $\alpha$ and $^7$Li escapes for detection by photomultiplier tubes. The average $^{10}$B film thickness does not exceed 300 nm. 



Below we first present the working principle of the detector and some relevant material properties, followed by some details of the detector design and construction. Next, we describe detector operation, detector performance, and correlate the detector performance with $^{10}$B-film characterization. The losses of efficiency are discussed towards the end, leaving room for further efficiency improvement through better understanding of the surface texture.


\section{Detection principle and material properties}
\label{princ:1}
\begin{figure}[thbp] 
   \centering
   \includegraphics[width=2.5in]{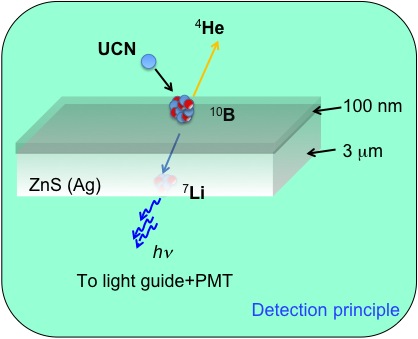} 
   \caption{The multilayer $^{10}$B surface detector for UCNs consists of a thin $^{10}$B top layer supported by a luminescent layer of ZnS:Ag. At least one of the charged particles $\alpha$ or $^7$Li generated from the neutron capture slows down or stops in the ZnS:Ag layer and emits light. A light-guide or a transparent window is used to transmit the light to a photomultiplier tube (PMT). A $^{10}$B thickness of 100 nm and a ZnS:Ag thickness of a few microns are sufficient.}
   \label{fig:PA}
\end{figure}

The working principle of the detector is illustrated in Fig.~\ref{fig:PA}. The detector design takes several lengths into account: UCN capture length ($\lambda_a$), $\alpha$ and $^7$Li ion ranges ($R^i$) in solid $^{10}$B and ZnS, and light attenuation in ZnS and light-guide. Compared with gas-based detectors, using only solid components in the detector removes the need for a material window. Compared with bulk $^7$Li- or $^{10}$B-doped scintillators for thermal neutrons, a 100-nm thick thin-film coating is sufficient since $\lambda_a$ is only 40 nm for UCNs at 4.4 m/s. An ideal UCN detection efficiency up to 95\% is expected for a film thickness of 3$\lambda_a$, or about 120 nm. When the UCN reflection from the $^{10}$B coated ZnS:Ag surface is taken into account, the efficiency can be reduced further by more than 20\% due to reflection, as shown below.

The ion ranges in $^{10}$B and ZnS are calculated using the Stopping and Range of Ions in Matter (SRIM) code~\cite{SRIM} and summarized in Table~\ref{Tb:range1}. Since the ion ranges are many times the $^{10}$B film thickness $\sim 3\lambda_a$, the charged particle energy losses in the $^{10}$B are small, except for ions that move at large angles with respect to the surface normal. For the 0.84 MeV $^7$Li, the full ion stopping in $^{10}$B only occurs when the angle is greater than $\theta_c = \cos^{-1} (3\lambda_a/R^i)$, or about 86 degrees for $\lambda_a=40$ nm. The corresponding loss of detection efficiency is about 3\% due to the 0.84 MeV $^7$Li loss alone. The total efficiency loss for the two branching ratios and both $\alpha$ and $^7$Li in the $^{10}$B layer is 
\begin{equation}
\epsilon_{loss} (^{10}{\rm B})= \sum_iw^i \frac{T_0}{R^i}
\end{equation}
for a flat uniform $^{10}$B layer thickness $T_0$. The values of $w^i$'s are given in Table~\ref{Tb:range1}. For $T_0 = 120$ nm, $\epsilon_{loss} (^{10}{\rm B}) = 5$\%. 



\begin{table}[htdp]
\caption{Maximum ion ranges ($R^i$) of the charged products from the $^{10}$B(n, $\alpha$)$^{7}$Li neutron capture process in $^{10}$B solid films and ZnS.}
\begin{center}
{\renewcommand{\arraystretch}{.60}
\renewcommand{\tabcolsep}{0.5 cm}
\begin{tabular}{lccc}
\hline
Ion & Energy & Range in  $^{10}$B & Range in  ZnS \\
(probability, $w^i$)&($E_0^i$, MeV)&($R^i$, $\upmu$m) & ($R^i$, $\upmu$m) \\\hline\hline
$\alpha$ (47\%)& 1.47 &  3.5 & 4.2 \\
$\alpha$ (3\%)& 1.78 &  4.4 & 5.1 \\
${}^{7}$Li (47\%)& 0.84 & 1.8 &2.3 \\
${}^7$Li (3\%) &1.02& 2.1 & 2.5  \\\hline
\end{tabular}}
\end{center}

\label{Tb:range1}
\end{table}%

\begin{figure}[htbp] 
 \centering
 \includegraphics[width=4in]{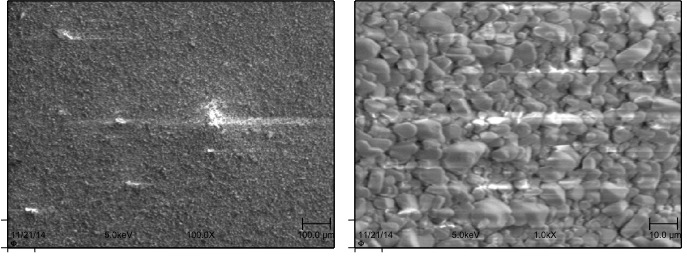} 
   \caption{(left) The 100-{\textmu}m resolution image of a $^{10}$B-coated ZnS surface using a scanning electron microscope (SEM). (Right) The 10-{\textmu}m resolution of the same film. The bright spots in both images are due to electrostatic charging of the surface.}
   \label{fig:SEM}
\end{figure}

ZnS:Ag coated acrylic acetate sheets (around 120 {\textmu}m thick) were obtained commercially~\cite{Peter:Jameson} and used as the substrates for $^{10}$B thin-film coating. According to the vendor, a transparent thermo-setting adhesive is applied to the acetate surface for ZnS:Ag bonding, so the ZnS:Ag facing the $^{10}$B is not coated with any adhesive. The ZnS:Ag powder is 16 {\textmu}m on average. Scanning electron microscope measurements of a lightly $^{10}$B-coated ZnS, Fig.~\ref{fig:SEM}, indicates the size dispersion of the ZnS powder, which also determines the surface roughness of the detector. 

Undoped ZnS emits light at 333 nm and 349 nm wavelengths, corresponding to the near bandgap energies at 3.55 eV and 3.72 eV respectively~\cite{Kroger:1954,Xiong:2004}. The near-bandgap emissions are strongly self-absorbed. ZnS can be doped in a variety of ways that shift the emission peaks to longer wavelengths and significantly reduce self-absorption~\cite{Feldman:2003,Li:2006}. The silver-doped ZnS emits blue light that peaks around 450 nm, with a characteristic decay time around 200 ns.
According to Knoll~\cite{Knoll:2000}, the relative light yield in ZnS:Ag due to $\alpha$ stopping is about 130\% of NaI(Tl) (3.8$\times$10$^4$ Photons/MeV), {\it i.e.}, the $\alpha$ light yield is 4.9$\times$ 10$^4$ photons/MeV. Based on the data sheets from Eljen Technology as well as Leo~\cite{Leo:1994}, the light yield of $\alpha$ is 300\% of anthracene (1.74$\times$10$^4$ photons/MeV), or 5.2$\times$10$^4$ photons/MeV. Their reports agree within 10\%. The absolute light yield for the present work is not quantified; so we use the previous data to estimate the light yield for the two different $\alpha$'s to be 7.4 $\times$ 10$^4$ photons (1.47 MeV) and 9.0 $\times$ 10$^4$ photons (1.78 MeV) respectively.

\section{Detector design and construction}


Highly enriched elemental $^{10}$B ($\sim$ 99 wt\%) in powder form was obtained from Ceradyne Boron Products. The $^{10}$B powder was placed in a carbon crucible and an electron-beam melted and evaporated the $^{10}$B onto the ZnS screens.  The distance between the crucible and the ZnS screen was about half a meter, sufficient to maintain the acrylic acetate sheets below 50$^o$C without any active cooling. A quartz microbalance (in-situ) and a small sapphire witness plate (offline) were used together to monitor the coating thickness. 
 
\begin{figure}[thbp] 
   \centering
   \includegraphics[width=4in]{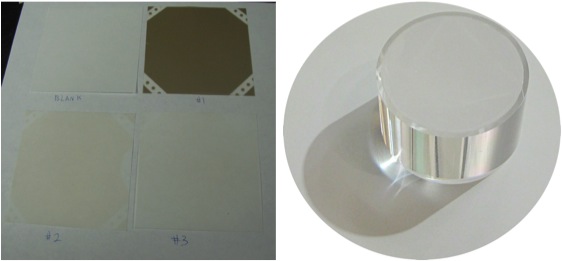} 
   \caption{(Left) Several $^{10}$B-coated ZnS:Ag screens with different coating thicknesses: the upper-right, the lower-left and the lower-right ones are compared with an uncoated blank screen (upper left). All of the screens are 10 cm $\times$ 10 cm in size. (Right) One of the $^{10}$B-coated ZnS:Ag screens is trimmed to a circle and glued to a 7.9 cm ( 3-1/8$''$) diameter acrylic lightguide (5 cm tall).}
   \label{fig:1}
 \end{figure}

 A few examples of the $^{10}$B-coated ZnS:Ag screens are shown in Fig.~\ref{fig:1}a. The darkest color one (upper right) corresponds to a thick $^{10}$B coating exceeding 200 nm. The thinner coatings are shown on the lower left and lower right. A blank ZnS:Ag screen is shown on the upper left for comparison.
 
 Two types of detectors have been built. In Fig.~\ref{fig:1}b, one of the lightly coated screens is cut to fit within the diameter of an acrylic lightguide. The screen is glued (Dow Corning Sylgard 184 Silicone Elastomer) to the lightguide. To remove the trapped air in the glue, the freshly glued screen is vacuum pumped down at a rate of 0.6 l/s for one to two  hours, until no visible bubbles exist in the silicone layer.
 
\begin{figure}[thbp] 
   \centering
   \includegraphics[width=3in]{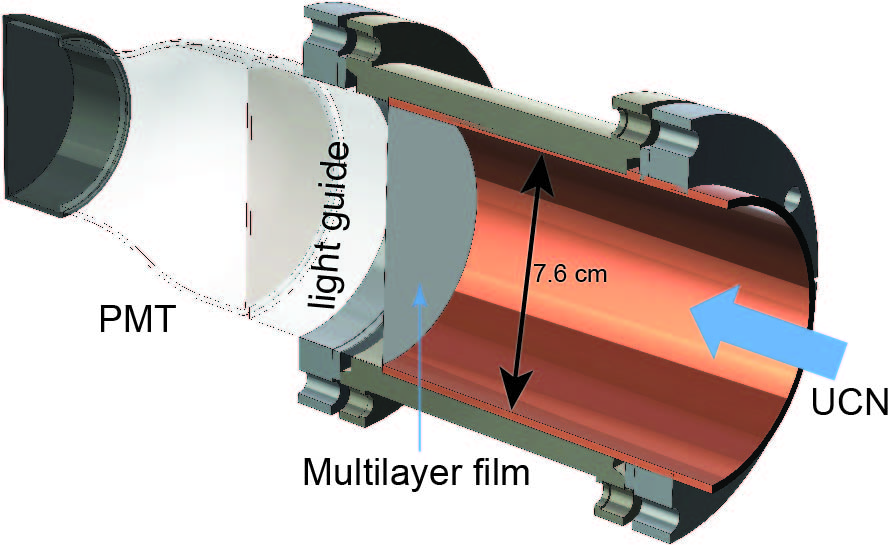} 
   \caption{A full detector assembly consists of the part shown in Fig.~\ref{fig:1}b and a PMT. A coupler connects the detector to a UCN guide with an inner diameter of 7.6 cm.}
   \label{fig:Ramsey}
\end{figure}

 A 7.6-cm (3$''$) diameter photomultiplier (PMT) is coupled to the ZnS-screen-covered  light-guide to form a detector as shown in Fig.~\ref{fig:Ramsey}. A rubber O-ring about 2.5 cm from the ZnS screen provides the vacuum seal. Half of the lightguide is inside the vacuum, the other half is outside the vacuum. This type of detector has been used for UCN flux monitoring, as well as flux studies as a function of height relative to the UCN beamline.

\begin{figure}[thbp] 
   \centering
   \includegraphics[width=3in]{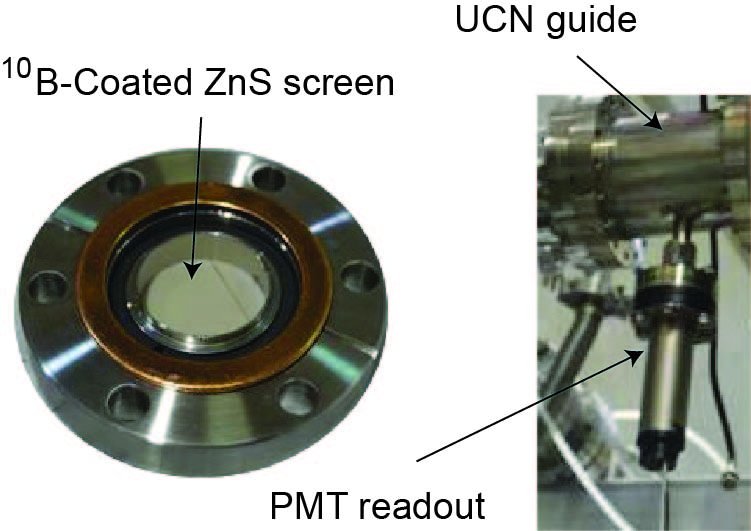} 
   \caption{(Left) A $^{10}$B-coated ZnS:Ag screen sits on the vacuum side of a 7-cm conflat flange; (Right) A PMT is attached to the air side of the flange. The full detector assembly is mounted to a UCN guide for flux monitoring. }
   \label{fig:Mor1}
 \end{figure}

A second type of detector with a smaller $^{10}$B area and without an acrylic lightguide is shown in Fig.~\ref{fig:Mor1} (Left). The 7-cm (2-3/4$''$) diameter conflat flange with a transparent window is used to house a $^{10}$B coated ZnS screen. A smaller (2.5 cm in diameter) PMT is then attached to the window from outside the vacuum. The whole detector assemble is attached to a port on a UCN guide, shown in Fig.~\ref{fig:Mor1} (Right). This type of detector was used to examine UCN detection efficiency, as well as for UCN transport studies in UCN guides and flux monitoring.

\section{Results and discussion}
\label{sec:DC}
Since there is ample light due to the $\alpha$ and $^7$Li ion stopping in ZnS:Ag, we could directly digitize the electric pulses from the PMT's. When an ORTEC 113 scintillation preamplifier was used (set to 0 input capacitance), we could reduce the PMT DC bias by up to 600 volts from -1.6 to -1.8 kV to -1.0 to -1.2 kV.  Most of the data were taken using an FADC-based waveform digitizer described in a previous work~\cite{Wang:2013}. 

\subsection{Pulse height spectra}
\begin{figure}[htbp] 
 \centering
 \includegraphics[width=3.5in,angle=0]{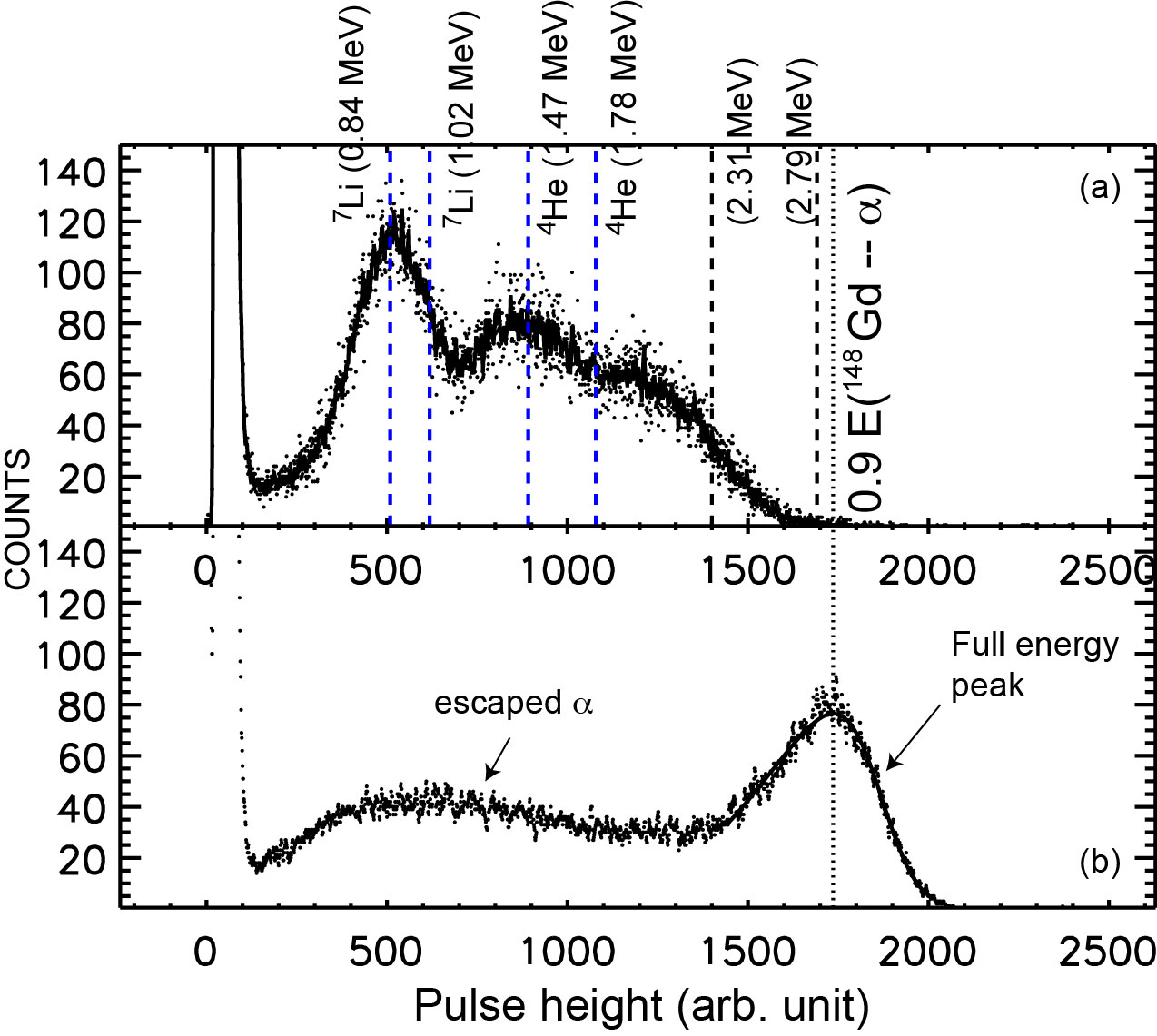} 
   \caption{(a) A typical pulse height spectrum due to UCN captures. (b) Energy calibration using a $^{148}$Gd $\alpha$ (3.182 MeV) source. The characteristic energies for $^7$Li and $\alpha$ due to UCN captures are shown in (a). The calibration line corresponds to 0.9 time the  $^{148}$Gd $\alpha$ energy. The $^{148}$Gd $\alpha $-peak is fitted with a skewed gaussian function.}
   \label{fig:PHS1}
\end{figure}

A typical pulse height spectrum (PHS) of a $^{10}$B-coated ZnS:Ag UCN detector with a lightguide is shown in Fig.~\ref{fig:PHS1}a. The energy scale was calibrated using a $^{148}$Gd $\alpha$ source (3.182 MeV). The PMT bias was at -1.7 kV. No ORTEC 113 scintillator preamp was used and the PMT output was digitized directly. The $^7$Li (0.84 MeV) peak stands out because the ion has the shortest range and the width of the peak is narrow due to the small straggling. A significant fraction of the signals appears above the 1.78 MeV $\alpha$ (only about 3\% is expected). This is attributed to the surface roughness as discussed in Fig.~\ref{fig:SEM} above. The average $^{10}$B coating (film $\#$3, this number is based on the order of film deposition) was measured to be 3.40 $\pm$ 0.04 nm using an optical profilometer. Since the surface roughness is greater than the $^{10}$B coating thickness, as well as the stopping ranges of $\alpha$ and $^7$Li in ZnS, both charge particles can be stopped in ZnS simultaneously if the charge particles leave the surface at an angle. 

For energy calibration, the $^{148}$Gd $\alpha$ peak was fit by a skewed Gaussian function~\cite{OhLe:1976},
\begin{equation}
f(x) = 2 c_1 \phi(x') \Phi(\alpha_s x') +c_0,
\end{equation}
with
\begin{equation}
x' = \frac{x-x_0}{\sigma},
\end{equation}
\begin{equation}
\phi(x) = \frac{1}{\sqrt{2\pi}} e^{-x^2/2},
\end{equation}
and \begin{equation}
\Phi(x) = \frac{1}{2} \left[1 +{\rm erf}(\frac{x}{\sqrt{2}})\right].
\end{equation}
The fitting parameters for the 3.182 MeV $^{148}$Gd $\alpha$ are $x_0 = 1867$, $\sigma = 299.0$, $\alpha_s=-3.636$, $c_1 = 110.2$, $c_0 =  1.104$. The same skewed Gaussian function can also be used to fit the 0.84 MeV $^7$Li peak. The fitting parameters for 0.84 MeV $^7$Li are  $x_0 = 430.0$, $\sigma = 162.5$, $\alpha_s=1.494$, $c_1 = 174.5$, $c_0 =  17.54$. The fact that the skewness parameter $\alpha_s$ is positive indicates that the 0.84 MeV $^7$Li pulse shape is distorted by the nearby 1.02 MeV $^7$Li and 1.47 MeV $\alpha$. 

\begin{figure}[htbp] 
 \centering
 \includegraphics[width=3in,angle=90]{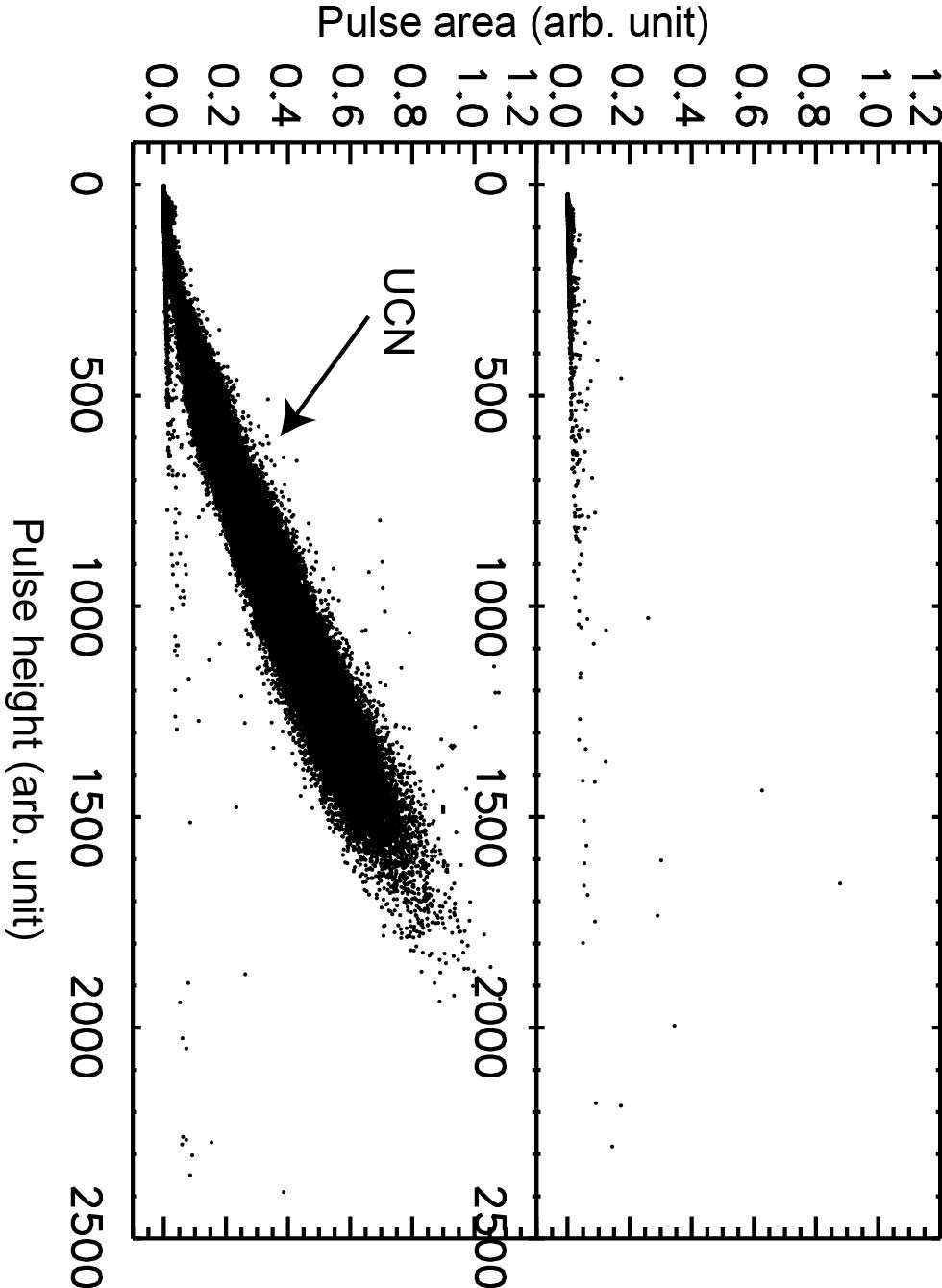} 
   \caption{(Top) Background signals when the gate valve to the detector is closed; (Bottom) UCN data with background as in Fig.~\ref{fig:PHS1}a. The pulse shape measurements confirm very low neutron induced background ($< 0.1$ Hz for the $^{10}$B surface area of 39.2 cm$^2$) in the detector.}
   \label{fig:PSD}
\end{figure}

The UCN-induced signals are also separated from most of the lower-amplitude background to the left ($< $ 200 on the horizontal scale), which is attributed to the light leaks from the UCN guide side. When the detector is gated off to the UCN guide, the UCN spectrum disappears and the low-amplitude background is also reduced significantly, as shown in Fig.~\ref{fig:PSD}. 


\subsection{Film thickness and efficiency}
\begin{figure}[htbp] 
 \centering
 \includegraphics[width=3in,angle=90]{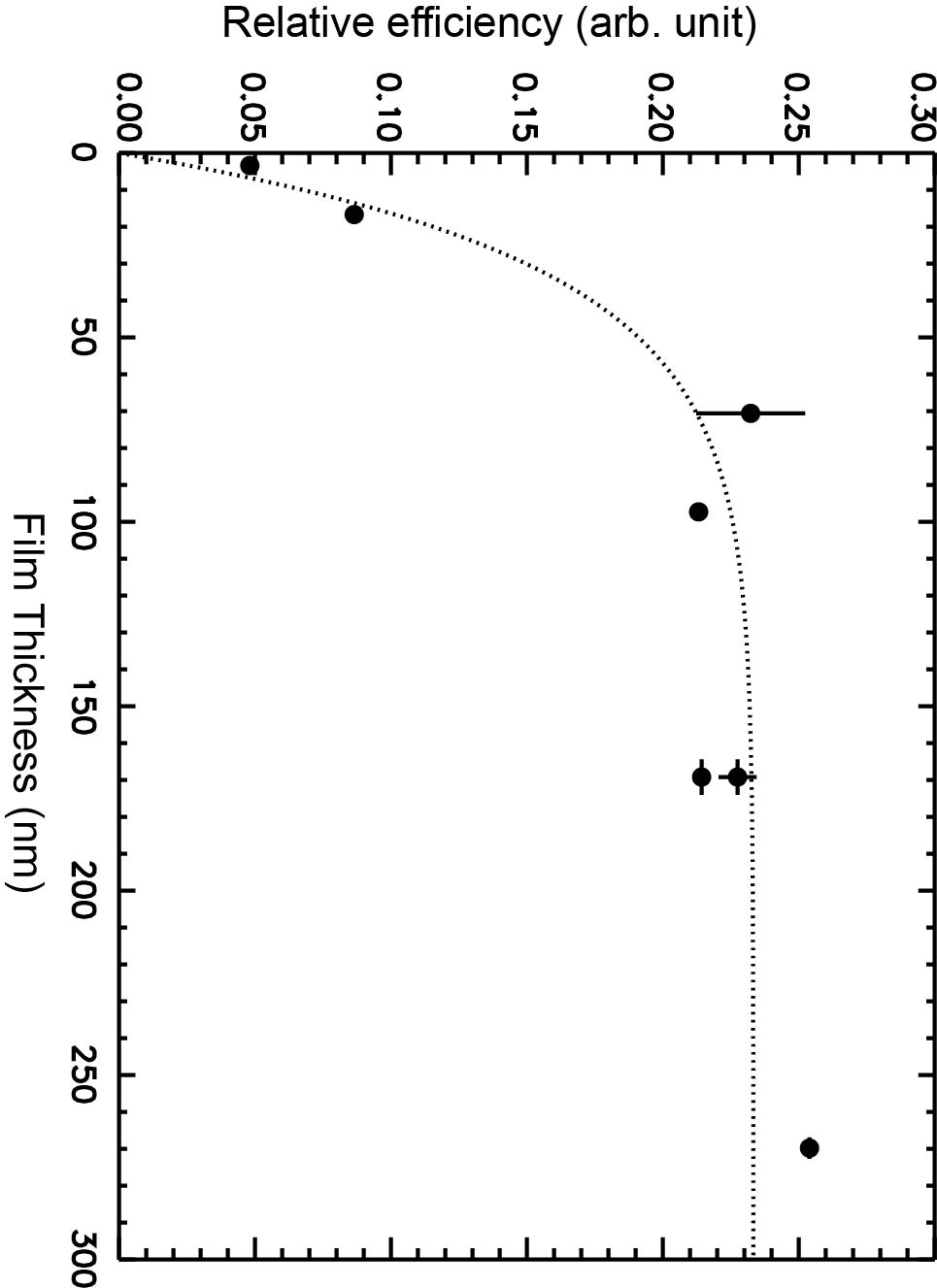} 
   \caption{Relative UCN efficiency as a function of $^{10}$B film thickness. The dash line is the fitting curve described by Eq.~(\ref{model:thick}).}.
   \label{fig:thickscan}
\end{figure}

We used the second detector configuration, similar to Fig.~\ref{fig:Mor1}, to characterize the UCN detection efficiency as a function of $^{10}$B film thickness. A $^3$He gas detector~\cite{Morris:1999} was used to normalize the film signals. Since the film areas were irregular, we placed a TPX screen (Polymethylpentene, Mitsui Chemicals) with a small hole ($\sim$ 6 mm diameter) in front of each film to limit the UCN flux to the same film area. UCNs that missed the hole were absorbed or upscattered to higher energies. The normalized UCN efficiency as a function of the average $^{10}$B film thickness is summarized in Fig.~\ref{fig:thickscan}. The film thickness was measured using a Zygo 3D optical surface profiler (NewView 7300). Because of the roughness of the ZnS surfaces, we used films deposited on sapphire witness plates (wafer grade) for thickness measurements. 
The data are fitted by
\begin{equation}
\epsilon(x) = a \left(1 - e^{-x/b}\right),
\label{model:thick}
\end{equation}
with $a = 0.23$ and $b = 29.3$ nm. Here $\epsilon(x)$ and $x$ are normalized efficiency and the corresponding film thickness respectively. The $b$ value corresponds to a mean UCN velocity of 3.2 m/s according to Eq.~(\ref{eq:mfp}) or a  mean UCN wavelength of 124 nm according to Eq.~(\ref{wvleng}).

\subsection{Impurities in the $^{10}$B films}
\begin{figure}[htbp] 
  \centering
   \includegraphics[width=3.5in]{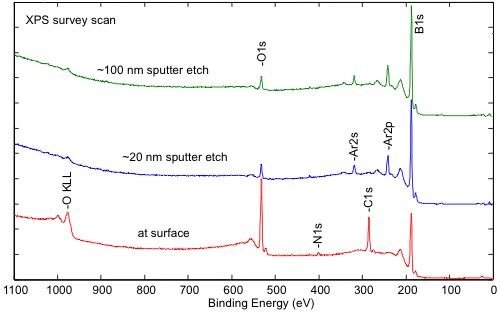} 
   \caption{X-ray photoelectron spectra at three different depths of a $^{10}$B film. }
   \label{fig:XPS}
\end{figure}

E-beam deposition took place in an ambient pressure $>10^{-6}$ Torr, which could introduce impurities to the $^{10}$B films through physical (adsorption) and chemical (4B+3O$_2 \rightarrow$ 2B$_2$O$_3$) processes. Ar ion sputtering and X-ray photoelectron spectroscopy (XPS) were combined to measure the impurity concentrations as a function of film depth. Some Ar contamination was observed from the XPS spectra, Fig.~\ref{fig:XPS}. The film (\#11) has a thickness of 269.8$\pm$2.9 nm from the optical profilometer measurement. 
The sputtering profile data indicates that a surface transition layer is $<$20 nm in thickness. XPS spectrum at $\sim$20 nm is similar to the XPS spectrum at 100 nm. The bulk of the thin film shows $>$ 95\% atomic B, $\sim$ 1\% atomic C, and about 3.5\% atomic O. The B$_2$O$_3$ layer is about 2.3 nm thick on the top of the B metal. The topmost hydrocarbon layer (paraffin equivalent) is about 0.5 nm thick. Little carbon contamination was observed for this sample. Higher carbon concentration ($> 10$\%) was observed in another sample, indicating that the crucible contamination can be controlled if the boron stock in the crucible is monitored closely during deposition.  

\subsection{Detection efficiency losses}
In addition to reduced detection efficiency due to charged particle stopping in the $^{10}$B layer, $\epsilon_{loss}$ = 5\% for 120-nm thick film as discussed above, other losses of detection efficiency come from coherent and incoherent scattering of UCN in the $^{10}$B layer. Multiple scattering effects such as reabsorption of scattered UCNs are neglected. We consider two limits. In the upper limit, we neglect the interference of coherently scattered UCNs. The UCN detection efficiency ($\epsilon_u$) is given by
\begin{equation}
\epsilon_u= \frac{f_0\sigma_{a0} n_0}{\sum_i f_i \sigma_{ti} n_i} \exp(- T \sum_i f_i \sigma_{ti} n_i),
\label{eq:m1}
\end{equation}
 for a film thickness of $T$. We use the subscript `$i$=0' for $^{10}$B, and $i\geq 1$ for other impurities such as $^{11}$B, $^{16}$O, {\it etc}. $f_i$ is the atomic fraction of the $i$th element and $n_i$ the corresponding number density. $n_0$ = $1.3 \times 10^{23} $ cm$^{-3}$. $\sigma_{a0}$ is the absorption cross section due to $^{10}$B. $\sigma_{ti}$ is the total cross section or the sum of absorption, coherent and incoherent scattering cross sections. In a pure $^{10}$B film, the ratio of the absorption cross section to the total cross section is 99.9\%. Therefore the scattering loss is insignificant compared to transmission loss. For sufficiently thick film $T \ge 3 \lambda_a$, above 95\% detection efficiency can be obtained. Impurities increase the scattered UCN loss. When the film is contaminated by 3.5\% of oxygen (mainly $^{16}$O with $\sigma_t$ = 4.232 barn) and 1\% of carbon ($\sigma_t$ = 5.551 barn), the ratio of $^{10}$B absorption to the total UCN attenuation is still 99.9\%.
  
In the lower detection limit, the interaction of UCNs with the absorbing surface is approximated by a complex Fermi potential $V_f ({\bf r}) = V({\bf r}) - iW({\bf r})$~\cite{Ignatovich:1990,Golub:1991}. Here {\bf r} indicates the position-dependence of the potential due to, for example, surface roughness and the position-dependent nuclear compositions of the surface. We follow the same sign convention for $W({\bf r})$ as in~\cite{Golub:1991}. For unpolarized UCNs interacting with an unpolarized surface, $V({\bf r}) = (2\pi\hbar^2/m) \sum_i n_i({\bf r}) b_i^{coh}$ is given by the number density of different nuclei $n_i({\bf r})$ and their corresponding bound coherent scattering length $b_i^{coh}$. 

Neglecting the surface roughness, a flat pure $^{10}$B surface has $V$ = -3.4 neV, $W$ = 36.3 neV, corresponding to the coherent scattering length of $b^{coh} = -0.1-1.066i $ fm for $^{10}$B~\cite{NIST}. When the film is contaminated by 3.5\% of oxygen (mainly $^{16}$O with $b^{coh} = 5.803$ fm) and 1\% of carbon (mainly $^{12}$C wtih $b^{coh} = 6.6511$ fm), the imaginary part of the Fermi potential $W$ remains the same (assuming that the $^{10}$B density remains the same, which is likely an over estimate). The real part of the Fermi potential becomes positive, $V =$ 5.8 neV. We can estimate the single-bounce UCN loss due to reflection for a smooth flat surface using the formula given in \cite{Lekner:1987},
\begin{equation}
R = \frac{(k_0 -k_r)^2 +k_i^2}{(k_0 + k_r)^2+k_i^2},
\label{reflectivity:1}
\end{equation}
where
$k_0 = \sqrt{2m E_0/\hbar^2} \cos \theta$. $k_r$ is given by
\begin{equation}
k_r^2 =(m/\hbar^2) [ \sqrt{(E_0 \cos^2\theta -V)^2 + W^2} + (E_0 \cos^2\theta -V)],
\label{eq:kr}
\end{equation}
and $k_i k_r = mW/\hbar^2$. Here $E_0$ is the incident UCN energy, $m$ the neutron rest mass, and $\theta$ the neutron incident angle with respect to the surface normal. By using Eq.~(\ref{reflectivity:1}), we have assumed that the $^{10}$B thickness is much greater than the neutron absorption length, which is valid for film thickness $\gtrsim $ 100 nm. Otherwise, $R$ needs to be modified into a form that is algebraically more complex, which can also be found in \cite{Lekner:1987}.

\begin{figure}[htbp] 
  \centering
   \includegraphics[width=3.5in,angle=0]{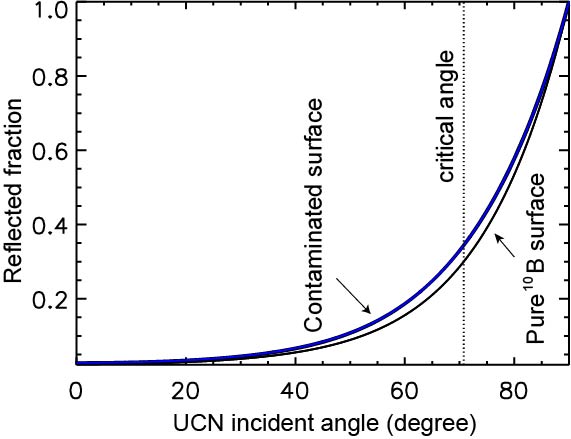} 
   \caption{UCN reflectivity as a function of incident angle on a flat surface. The mean UCN energy of 53.5 neV is used. The critical angle for total reflection is 70.8 degrees, corresponding to $V$ = 5.8 neV.}
   \label{fig:Reflect}
\end{figure}

The reflected UCN fraction as a function of incident angle is plotted in Fig.~\ref{fig:Reflect} for the mean UCN energy of 53.5 neV. The critical angle, as in the cases of non-absorbing surface, or $W = 0$, is given by $\theta_c = \cos^{-1} (\sqrt{V/E_0})$, or 70.8 degrees. For non-absorbing surfaces, $R=1$ at incident angles greater than the critical angle, $\theta > \theta_c$. For the highly absorbing $^{10}$B surface, we notice that $R < 1$ for $\theta > \theta_c$. In Eq.~(\ref{eq:kr}), $k_r$ does not vanish due to finite $W$. Because of surface roughness, however, the incident angle is not well defined. The mean reflectivity averaged over the incident angles is $\int_0^{\pi/2} R \sin \theta d\theta$. The averaged UCN reflectivity for a pure $^{10}$B surface is 26.5\%. For a slightly contaminated surface it is 29.0\%. This relatively small difference between a pure $^{10}$B surface and a slightly contaminated surface is mainly due to the reduction in $R$ for $\theta > \theta_c$. The measurements in Fig.~\ref{fig:thickscan} indicate that the UCN detection efficiency is rather insensitive to impurities of a few percent, consistent with Fig.~\ref{fig:Reflect}. The discrepancy in efficiency loss between the single-particle scattering model Eq.~(\ref{eq:m1}) and the optical model will be studied by examining the effects of surface roughness. 


\section{Conclusions}

We have demonstrated a multilayer surface detector for ultracold neutrons (UCNs). No gas is used in the detector and the UCNs are directly captured by a $^{10}$B surface. The  spectra measured from UCN captures on the $^{10}$B-films consist of primarily 0.84 MeV $^7$Li and 1.47 MeV $\alpha$, as well as a significant fraction of signals which are attributed to dual-charge ($^7$Li and $\alpha$ together) stopping due to the uneven ZnS surface. Ambient neutron background count rate was observed to be less than 0.1 Hz (for the $^{10}$B surface area of 39.2 cm$^2$ and the film thickness of 3.4 nm) using pulse-shape discrimination.  This type of detector has been configured in several ways for UCN flux monitoring, development of UCN guides, UCN lifetime measurement~\cite{Salvat:2014} and UCN-induced fission research~\cite{Broussard:2014}. Further work is needed to understand the effects of surface roughness on the detection efficiency. 

{\bf Acknowledgments} This work was funded by the LDRD program of Los Alamos National Laboratory. 

\bibliographystyle{elsarticle-num}
\bibliography{ZnSB10v4}







\end{document}